\documentclass[letterpaper]{article} 
\usepackage[preprint]{aaai2027}  
\usepackage[hyphens]{url}  
\usepackage{graphicx} 
\usepackage{natbib}  
\usepackage{caption} 
\usepackage{algorithm}
\usepackage{algorithmic}
\usepackage{amsmath}
\usepackage{newfloat}
\usepackage{listings}
\DeclareCaptionStyle{ruled}{labelfont=normalfont,labelsep=colon,strut=off} 
\floatstyle{ruled}
\newfloat{listing}{tb}{lst}{}
\floatname{listing}{Listing}

\usepackage{booktabs}

\usepackage{tcolorbox}
\tcbuselibrary{breakable,listings,skins}

\definecolor{promptblue}{RGB}{20,32,170}
\definecolor{promptgray}{RGB}{248,248,248}

\newtcblisting[
    auto counter,
    number within=section
]{promptbox}[2][]{%
    enhanced,
    breakable,
    listing only,
    title={Prompt~\thetcbcounter: #2},
    colback=promptgray,
    colframe=promptblue,
    colbacktitle=promptblue,
    coltitle=white,
    fonttitle=\bfseries,
    boxrule=0.7pt,
    arc=1.5mm,
    outer arc=1.5mm,
    left=1.5mm,
    right=1.5mm,
    top=1.2mm,
    bottom=1.2mm,
    before skip=6pt,
    after skip=6pt,
    listing options={
        basicstyle=\ttfamily\footnotesize,
        breaklines=true,
        breakatwhitespace=false,
        columns=fullflexible,
        keepspaces=true,
        showstringspaces=false,
        tabsize=2
    },
    #1
}

\title{Lineage-Aware Search Control for LLM-Based Evolutionary Factor Mining}
\author{
    Xintong Liu,
    Chenkai Zhang,
    Yanzhao Ma,
    Peng Yang\corresponding
}

\affiliations{
    Southern University of Science and Technology (SUSTech)\\
    Shenzhen, China\\
    12311926@mail.sustech.edu.cn
}

\begin{document}

\maketitle

\begin{abstract}
Formulaic alpha factor mining searches for symbolic expressions
that transform market data into predictive signals. Recent
LLM-based evolutionary factor mining methods use large language
models as semantic variation operators. Under the conventional
LLM-EA paradigm, stochastic variation and fitness-based selection
can leave the search weakly directed and concentrated around a
narrow set of structurally similar factors. We argue that these limitations
arise because current methods mainly
organize search at the individual and population levels, leaving
lineage-level control underused. A lineage connects successive
factor variants through evolutionary descent. We propose \textsc{LineageEvo}, which uses lineage in two coupled forms of search
control: (1) lineage-conditioned LLM variation, which distills Lineage-Scoped Search Experience into operator-specific
priors and fuses them with global operator priors, providing
persistent, lineage-relevant guidance for LLM generation; and
(2) lineage-stratified search allocation, which balances factor
quality against lineage concentration by distributing the search
budget across lineages during evolutionary search. Together,
these mechanisms make semantic variation more direction
aware and prevent branches from dominating the evolution,
thereby promoting the discovery of predictive and diverse factors. We evaluate \textsc{LineageEvo} on CSI 500 and S\&P 500
in terms of factor predictiveness, portfolio performance, search
efficiency, and lineage- and expression-level diversity.
\textsc{LineageEvo} achieves the strongest predictive
performance on both markets, while LSPS produces a more balanced lineage distribution
and a structurally broader factor pool. The results support
lineage as a useful intermediate unit for controlling LLM-based
evolutionary factor search.
\end{abstract}


\section{Introduction}

Formulaic alpha factor mining aims to discover mathematical expressions that transform raw market data into predictive signals for future asset returns \citep{formulaic101}. Such factors support portfolio construction
and trading, yet the vast expression space and noisy,
non-stationary data make predictive and stable factors difficult
to discover.

Early factor discovery largely relied on human experts who designed signals based on financial intuition and empirical experience. To reduce such manual effort, automated factor mining methods have been developed, including genetic programming (GP), reinforcement learning (RL), and other machine learning approaches \citep{autoalpha,alphagen,alphasage}. These methods expand the searchable factor space and enable systematic exploration beyond hand-crafted factors. However, they still rely
mainly on fixed grammars, handcrafted operators, or scalar
rewards, limiting semantic reasoning and adaptive reuse of search
experience.

More recently, LLM-based factor mining has emerged as a
promising paradigm, using pretrained models to generate and
refine symbolic formulas through agent-based generation, tree
search, and graph-aware evolution
\citep{alphagpt,alphaagent,llm_mcts_alpha,alphaprobe,alphabench}.
Large-scale pretraining provides semantic priors over financial
concepts, mathematical expressions, and programming patterns,
supporting factor hypothesis generation and formula refinement.
Among these approaches, LLM-based evolutionary algorithm
(LLM-EA) factor mining is particularly appealing because it
replaces purely syntactic genetic operators with LLM-based
semantic mutation and recombination
\citep{lmea,evoalpha,treevo,quantaalpha}. However, two
search-control challenges remain. First, LLM-based operators are flexible but weakly controllable, as the mapping from prompt context to offspring is implicit and difficult to steer persistently \citep{lmea,llm_mutation_control}. Second, under limited budgets, fitness-based selection can lead to takeover, where currently strong individuals or branches receive disproportionate search opportunities, reducing diversity and risking premature convergence \citep{sudholt2018diversity,gabor2018inheritance}.

These challenges persist because current LLM-EA factor mining
methods mainly organize search at the individual and population
levels, while making limited use of lineage-level state. At the
individual level, LLM operators modify selected parent factors;
at the population level, evolutionary selection favors
high-scoring candidates. Between them, a \emph{lineage} is a
descent-defined factor family whose membership follows primary
evolutionary inheritance. Unlike an isolated individual or an unstructured population, a lineage preserves the continuity of local search across successive variation events, making it a natural stateful unit for both persistent operator guidance and search allocation. Conventional
evolutionary algorithms have used genealogy and lineage
information for diversity preservation, selection, and
inheritance analysis \citep{burke2003lineage,gabor2018inheritance,burlacu2013lineages}.
However, using lineage as a shared scope for operator priors
and search allocation remains underexplored in LLM-EA factor mining.

To address these issues, we propose \textsc{LineageEvo}, a
lineage-aware LLM-EA framework for formulaic factor mining.
\textsc{LineageEvo} represents evolutionary descent in a lineage
DAG and uses lineage in two coupled forms of search control.
First, it organizes Lineage-Scoped Search Experience (LSSE) into
operator-specific lineage priors and fuses them with global
operator priors to condition LLM-based mutation and crossover.
Second, Lineage-Stratified Parent Selection (LSPS) allocates
search opportunities at the lineage level before selecting
parent factor(s), while a separate elite archive preserves
high-quality discoveries. This design gives LLM variation a persistent local context and
distributes subsequent search more evenly across evolutionary
branches.

Experiments on CSI 500 and S\&P 500 demonstrate the effectiveness
of \textsc{LineageEvo} across markets. Compared with the strongest
baseline for each metric, \textsc{LineageEvo} improves IC by
28.3\% and 12.6\%, and ICIR by 9.9\% and 35.8\% on CSI 500 and
S\&P 500, respectively, while achieving the strongest portfolio performance on both
markets. Ablation results
confirm the complementary roles of lineage-specific and global
priors, while LSPS increases lineage entropy from 0.405 to 2.154
and average AST distance from 0.509 to 0.747, preserving a broader
and less concentrated factor pool. The search curve further shows
that \textsc{LineageEvo} reaches a strong factor set rapidly,
indicating efficient evolutionary search.

Our contributions are summarized as follows:
\begin{itemize}

\item We identify lineage-level search control as underexplored in LLM-EA factor mining and use it to connect
individual-level semantic variation with population-level
selection.

\item We propose \textsc{LineageEvo}, which represents
descent-defined factor families in an evolutionary DAG and
organizes Lineage-Scoped Search Experience into
operator-specific lineage priors for LLM variation.

\item We introduce lineage-stratified search allocation through
LSPS, which balances lineage quality against lineage concentration, and decouple search allocation from solution
preservation through a separate elite archive.

\item Experiments on CSI 500 and S\&P 500 show that
\textsc{LineageEvo} outperforms eight representative baselines
in factor performance, while preserving greater lineage
and structural diversity.
\end{itemize}

\section{Related Work}

\subsection{Traditional Automated Factor Mining}

Traditional automated factor mining mainly formulates alpha discovery
as symbolic search. AutoAlpha applies GP-style variation and
quality-diversity search, AlphaGen uses reinforcement learning,
and AlphaSAGE improves structure-aware exploration
\citep{autoalpha,alphagen,alphasage}. These methods broaden
automated search, but still largely depend on predefined grammars, handcrafted operators, or scalar rewards, making it difficult to incorporate domain semantics or adaptively reuse search experience beyond predefined operators.

\subsection{LLM-Based Factor Mining}

Recent work introduces LLMs into factor mining as semantic generators, refiners, and search assistants. Alpha-GPT explores human--AI interactive alpha generation by translating investment ideas into candidate formulas \citep{alphagpt}, while AlphaAgent develops an autonomous agent framework with originality, hypothesis-factor alignment, and complexity control \citep{alphaagent}. Search-oriented methods further combine LLM generation with Monte Carlo Tree Search or graph-based retrieval to guide formula refinement \citep{llm_mcts_alpha,alphaprobe}. These methods make factor mining more semantic by leveraging LLMs' pretraining-induced knowledge of financial concepts, mathematical expressions, and programming patterns to generate factors with richer rationales and more flexible structures. However, as suggested by AlphaBench, existing LLM-based factor mining methods may still under-exploit the exploration capacity of LLMs, motivating evolutionary frameworks that embed LLMs into iterative population-based search \citep{alphabench}.

\subsection{LLM-EA Factor Mining}

LLM-EA search combines the semantic generation ability of LLMs with the population-based optimization mechanism of evolutionary algorithms. General LLM-EA studies use LLMs as evolutionary optimizers or hyper-heuristics, where LLMs generate, mutate, recombine, or reflect on candidate solutions within an iterative search loop \citep{lmea,eoh,reevo}. In factor mining, EvoAlpha adapts this paradigm to evolutionary alpha discovery with LLM-generated factors \citep{evoalpha}; TreEvo evolves tree-structured thoughts to capture the hierarchical nature of alpha formulas \citep{treevo}; and QuantaAlpha explores trajectory-level mutation and crossover to reuse effective mining experience \citep{quantaalpha}. These methods move beyond isolated LLM generation by introducing evolutionary variation and selection into factor search. However, they still face two fundamental search-control challenges: LLM-driven variation can be flexible but weakly controllable, while fitness-based selection can cause dominant branches to monopolize the search process.

\subsection{Genealogy and Lineage in Evolutionary Algorithms}

Evolutionary computation has previously used genealogy and
lineage information for diversity preservation, inheritance
analysis, and selection
\citep{burke2003lineage,gabor2018inheritance,
burlacu2013lineages}. These studies establish lineage
as an evolutionary abstraction. Building on this
foundation, \textsc{LineageEvo} uses each lineage as the shared
scope for accumulating operator-specific priors
and allocating subsequent search opportunities in LLM-EA factor
mining.

\begin{figure*}[t]
\centering
\includegraphics[
    width=0.95\textwidth,
    trim=3.3cm 6.4cm 3.52cm 3.63cm,
    clip
]{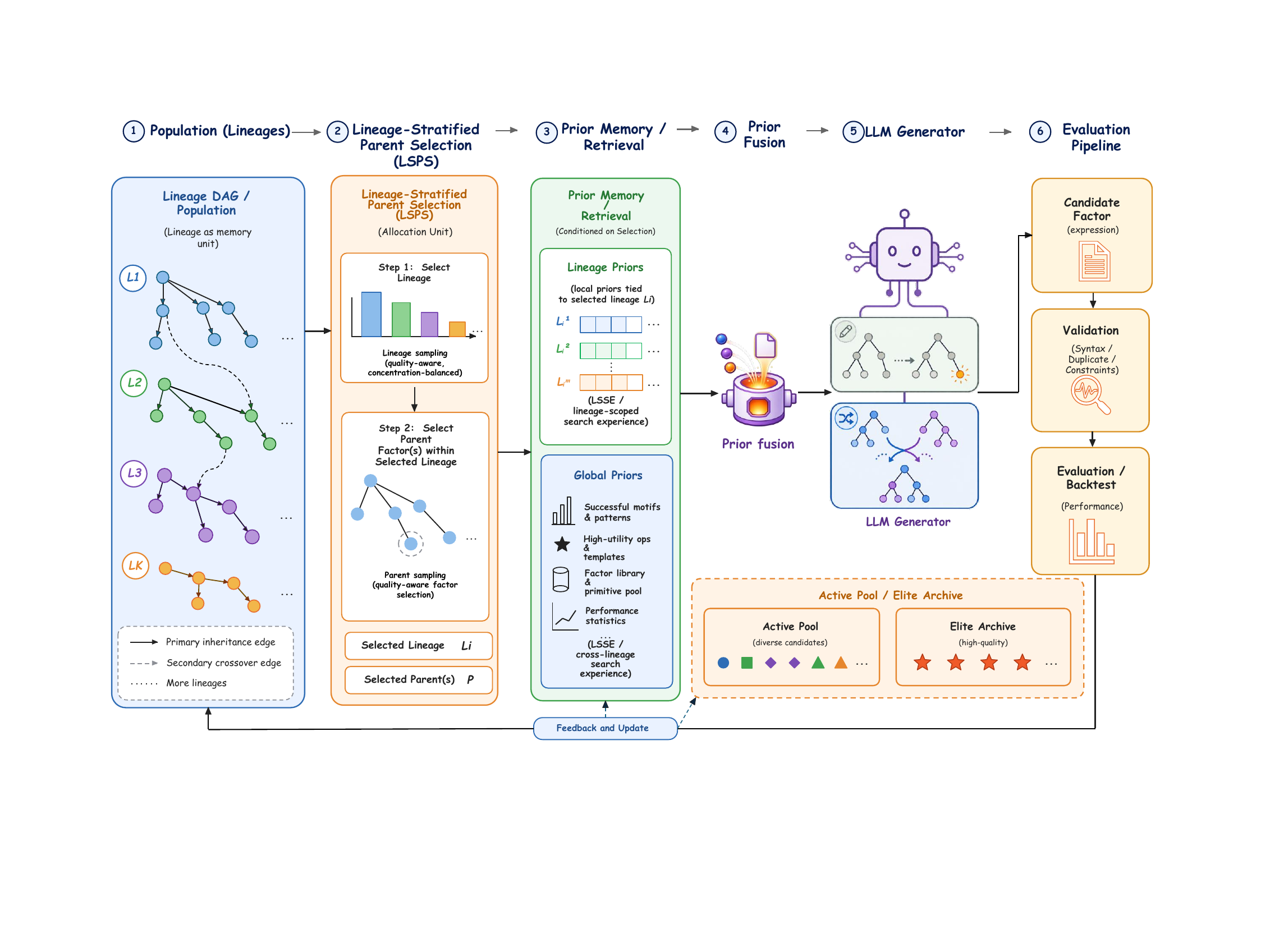}
\caption{
Overview of \textsc{LineageEvo}. LSPS
first selects a lineage and then selects parent factor(s);
lineage-specific and global operator priors are subsequently
retrieved and fused to condition LLM-based variation. Valid
offspring are evaluated, and the feedback path updates the
lineage DAG, prior states, active population, and elite archive.
}
\label{fig:overview}
\end{figure*}

\section{Method}
We propose \textsc{LineageEvo}, a lineage-aware evolutionary
framework for formulaic alpha factor mining with LLM-based
semantic variation. Its central idea is to associate accumulated search experience and
future search opportunities with each lineage.  \textsc{LineageEvo} uses lineage as a
stateful intermediate search unit between individual-level LLM
variation and population-level evolutionary selection.
Specifically, lineage serves two complementary roles: (1) it defines
a local scope for accumulating operator-specific search
experience, and (2) it provides a unit for allocating a
limited evaluation budget.

Figure~\ref{fig:overview} summarizes one iteration of
\textsc{LineageEvo}. Starting from an evolutionary state organized by lineage, Lineage-Stratified Parent Selection (LSPS) first allocates
a search opportunity to a lineage and then selects parent
factor(s) from the active population. The corresponding
lineage-specific and global operator experience is retrieved and
fused into soft guidance for LLM-based mutation or crossover. A
deterministically validated offspring is subsequently evaluated,
after which the lineage graph, prior states, active population,
and elite archive are updated. This division of responsibilities
keeps the LLM focused on semantic variation, while deterministic
components govern validation, evaluation, inheritance, and
search-state transitions.

\subsection{Problem Setup and Lineage Representation}

Let $\mathcal{X}$ denote the market feature space and
$\mathcal{F}$ the symbolic expression space induced by a factor
grammar. A formulaic alpha factor $f\in\mathcal{F}$ maps the
historical observations of stock $i$ to a cross-sectional signal:
\begin{equation}
    z_{i,t}^{f}
    =
    f\!\left(\mathcal{X}_{i,\leq t}\right),
    \label{eq:factor_definition}
\end{equation}
where $z_{i,t}^{f}$ is the factor value at date $t$. Rather than seeking a single best expression, the
goal is to discover a collection of factors that exhibit strong
predictive performance and temporal stability while maintaining
sufficient structural diversity during search. The resulting
factor pool supports downstream factor selection and combination.

\textsc{LineageEvo} ties lineage identity to the structural
inheritance induced by its variation operators. Mutation applies
a local modification to a leaf node or a small subtree of one
parent while preserving the remaining expression structure.
Crossover is asymmetric: the parent sampled from the first selected
lineage is fixed as the primary parent and provides the recipient
expression backbone. A subtree from the secondary parent replaces
a selected primary subtree, while the remaining primary structure
is preserved.

We record these inheritance relations using a directed acyclic
graph
\begin{equation}
    \mathcal{G}^{t}
    =
    \left(
        \mathcal{V}^{t},
        \mathcal{E}_{\mathrm{pri}}^{t},
        \mathcal{E}_{\mathrm{sec}}^{t}
    \right),
    \label{eq:lineage_dag}
\end{equation}
where $\mathcal{V}^{t}$ contains all valid factors evaluated by
iteration $t$. The set $\mathcal{E}_{\mathrm{pri}}^{t}$ records
the inheritance of the expression backbone. A mutation offspring
has one primary edge from its parent, whereas a crossover
offspring has one primary edge from the backbone parent. The set
$\mathcal{E}_{\mathrm{sec}}^{t}$ records the donor ancestry of
crossover, preserving the contribution of the parent that
provides the replacement subtree.

Let $\mathcal{R}^{t}\subseteq\mathcal{V}^{t}$ denote the set of
root factors. Since each non-root node has exactly one incoming
primary inheritance edge, its lineage root is recursively defined
as
\begin{equation}
    \rho^{t}(v)
    =
    \begin{cases}
        v,
        & v\in\mathcal{R}^{t}, \\[1mm]
        \rho^{t}(u),
        & (u,v)\in\mathcal{E}_{\mathrm{pri}}^{t}.
    \end{cases}
    \label{eq:lineage_root}
\end{equation}
The lineage associated with root $r_k$ is consequently
\begin{equation}
    \mathcal{L}_{k}^{t}
    =
    \left\{
        v\in\mathcal{V}^{t}
        \mid
        \rho^{t}(v)=r_k
    \right\}.
    \label{eq:lineage_definition}
\end{equation}

A lineage is therefore a descent-defined local search family
induced by structure-preserving variation. Lineage membership follows the
inherited expression backbone through primary edges, while
secondary edges retain cross-lineage donor ancestry. Because mutation modifies only a local component and crossover
preserves the primary backbone, nearby descendants are expected
to retain greater structural and DSL-level financial-semantic
continuity than factors from unrelated lineages, as verified in
Appendix. This continuity provides
the basis for using lineage as a scope for search experience and
resource allocation.

\subsection{Lineage-Stratified Search Allocation}

Modules 1--2 in Figure~\ref{fig:overview} implement
lineage-level search allocation. Standard individual-level
fitness selection may repeatedly favor factors from already
dominant branches, causing early lineage takeover and reducing
exploration
\citep{sudholt2018diversity,gabor2018inheritance}. To mitigate
this effect, \textsc{LineageEvo} organizes its active population
by lineage:
\begin{equation}
    \mathcal{P}^{t}
    =
    \bigcup_{k=1}^{m_t}
    \mathcal{P}_{k}^{t},
    \qquad
    \mathcal{P}_{k}^{t}
    =
    \mathcal{P}^{t}\cap\mathcal{L}_{k}^{t},
    \label{eq:lineage_population}
\end{equation}
where $\mathcal{P}_{k}^{t}$ contains the active factors belonging
to lineage $k$, and $m_t$ is the number of active lineages.

LSPS uses a two-stage sampling procedure. It first selects a
lineage and then selects a parent factor within that lineage.
Both stages use a common robust quality score. Let $\mathcal{M}$
denote the optimization metric, either IC or ICIR, and define
$T(f)=|\mathcal{M}_{\mathrm{tr}}(f)|$ and
$V(f)=|\mathcal{M}_{\mathrm{val}}(f)|$. For factors satisfying
$T(f)+V(f)>0$, the quality score is
\begin{equation}
    q(f)
    =
    \alpha T(f)
    +
    \beta V(f)
    -
    \gamma
    \left|
        \mathcal{M}_{\mathrm{tr}}(f)
        -
        \mathcal{M}_{\mathrm{val}}(f)
    \right|,
    \label{eq:parent_score}
\end{equation}
where $\alpha,\beta>0$ and
$0\leq\gamma<\min\{\alpha,\beta\}$. This guarantees
$q(f)>0$ while penalizing both strength disagreement and
directional reversals.

For each lineage, we define a quality score and a lineage-balance
weight:
\begin{equation}
    b_k^{t}
    =
    \max_{f\in\mathcal{P}_{k}^{t}}q(f),
    \qquad
    d_k^{t}
    =
    \frac{1}{|\mathcal{P}_{k}^{t}|+1}.
    \label{eq:lineage_weights}
\end{equation}
The first term reflects the strongest active factor within the
lineage, while the second assigns greater weight to less
represented lineages. Their normalized mixture defines the
lineage-selection distribution:
\begin{equation}
    \pi_{\mathrm{lin}}^{t}(k)
    =
    \lambda
    \frac{b_k^{t}}
    {\sum_{\ell=1}^{m_t}b_{\ell}^{t}}
    +
    (1-\lambda)
    \frac{d_k^{t}}
    {\sum_{\ell=1}^{m_t}d_{\ell}^{t}},
    \label{eq:lineage_sampling_probability}
\end{equation}
where $\lambda\in[0,1]$ controls the trade-off between favoring
high-quality lineages and maintaining exploration across
lineages.

After selecting lineage $k$, LSPS samples a parent in proportion
to its robust quality:
\begin{equation}
    \pi_{\mathrm{par}}^{t}(f\mid k)
    =
    \frac{q(f)}
    {\sum_{g\in\mathcal{P}_{k}^{t}}q(g)},
    \qquad
    f\in\mathcal{P}_{k}^{t}.
    \label{eq:parent_sampling_probability}
\end{equation}
This two-stage probabilistic selection exploits stronger factors
while preserving alternatives at both the lineage and parent
levels.

For mutation, one parent is sampled from the selected lineage.
For crossover, this first parent $p_t^{a}$ is fixed as the
primary parent, while a secondary donor $p_t^{b}$ is drawn from
another lineage whenever possible under the same sampling rule.
The allocated and inherited lineages therefore coincide:
\begin{equation}
    p_t^{\star}
    =
    p_t^{a},
    \qquad
    \rho^{t+1}(f_{t+1})
    =
    \rho^{t}(p_t^{\star})
    =
    r_k.
    \label{eq:primary_parent}
\end{equation}
The secondary parent contributes the donor subtree and is
retained through a secondary ancestry edge.

Search allocation and solution preservation are handled
separately. The lineage-bucketed active population governs future
parent selection, while an independent elite archive
$\mathcal{A}^{t}$ retains high-quality factors for final
selection.

\subsection{Lineage-Conditioned LLM Variation}

Modules 3--5 in Figure~\ref{fig:overview} use evolutionary
experience associated with each lineage to condition LLM-based
semantic variation. Each valid offspring produces an evolutionary
event that records the parent and offspring structures, their
expression difference, the applied operator, and train/validation
evidence. A mutation event is associated with its parent lineage,
whereas a crossover event is associated with the primary lineage
while retaining the secondary donor relation. We refer to the
experience accumulated within a lineage as
\emph{Lineage-Scoped Search Experience} (LSSE).

Mutation and crossover require different forms of experience.
Mutation priors summarize beneficial and harmful local
modifications within one lineage, whereas crossover priors capture
transferability, complementarity, and incompatibility between
recipient and donor structures. Accordingly, each lineage
maintains separate priors $S_{k,o}^{t}$ for
$o\in\{\mathrm{mut},\mathrm{cross}\}$, while $G_o^{t}$ stores the
corresponding cross-lineage experience. Mutation uses the prior
of its parent lineage, whereas crossover uses both parent-lineage
priors, with the primary prior describing the recipient backbone
and the secondary prior providing donor-side guidance.

The priors are updated incrementally by an event-triggered LLM
rewriter:
\begin{equation}
\begin{aligned}
    S_{k,o}^{t+1}
    &=
    \Psi_o
    \left(
        S_{k,o}^{t},
        e_{t+1}
    \right), \\
    G_o^{t+1}
    &=
    \Psi_o^{\mathrm{glob}}
    \left(
        G_o^{t},
        e_{t+1}
    \right).
\end{aligned}
\label{eq:prior_update}
\end{equation}
Here, $k$ denotes the mutation-parent lineage or the crossover
primary lineage. For crossover, only the primary-lineage prior and global crossover prior are
updated because the offspring inherits the primary backbone; the
secondary prior remains read-only. When an improvement or degradation trigger is activated, the
rewriter uses its financial and mathematical knowledge to
interpret how the parent--offspring structural change alters the
factor's signal construction and whether the performance metric change supports or contradicts the
existing guidance. It then revises the structured prior by
strengthening supported hypotheses, weakening contradicted ones,
and recording reusable transformation principles or failure
warnings. Mutation updates summarize beneficial and harmful local
edits, while crossover updates describe transferable donor
structures and their compatibility with the primary backbone.
A lightweight deterministic manager validates the structured
output and retains the previous state if rewriting fails.

We use the term \emph{prior} to denote a structured
pre-generation inductive bias rather than a normalized Bayesian
distribution. Distilled from previous evolutionary evidence and
provided before the next variation step, it biases the LLM toward
historically promising transformations.

Let $\mathbf{S}_t$ denote the lineage-prior input for the current
variation. The conditioning context is constructed as
\begin{equation}
    C_t
    =
    \Phi_{o_t}
    \left(
        \mathbf{S}_t,
        G_{o_t}^{t};
        \lambda_{k_t,o_t}^{t}
    \right),
    \label{eq:prior_fusion}
\end{equation}
where $\Phi_o$ is a deterministic fusion and prompt-rendering
policy. The emphasis parameter $\lambda_{k,o}^{t}$ is regulated
from the recent evolutionary state of lineage $k$, including its
performance trend and stagnation status. When the lineage
continues to improve, the policy places greater emphasis on its
lineage-specific prior to exploit locally effective variation.
Under stagnation or degradation, it increases the emphasis on
the global prior to introduce transferable cross-lineage
experience and reduce local search bias. The policy then fuses
the two prior sources and renders the resulting context as
concise natural-language guidance for the candidate LLM.

Let $P_t$ contain either one mutation parent or an ordered
primary--secondary crossover pair, and let $\Gamma$ denote the
factor grammar and generation constraints. The LLM induces
\begin{equation}
    f_{t+1}
    \sim
    p_{\theta}
    \left(
        \cdot
        \mid
        o_t,
        P_t,
        C_t,
        \Gamma
    \right).
    \label{eq:offspring_generation}
\end{equation}
The conditioning context acts as soft semantic guidance rather
than a hard search constraint. Consequently, the LLM retains the
flexibility to construct new factor structures while receiving persistent lineage-relevant context unavailable from the
current parents alone.
\subsection{Feedback and Search-State Update}

Module 6 and the feedback path in Figure~\ref{fig:overview}
complete each search iteration. We summarize the search state at
iteration $t$ as
\begin{equation}
    \Omega^{t}
    =
    \left(
        \mathcal{G}^{t},
        \mathcal{P}^{t},
        \mathcal{A}^{t},
        \{S_{k,o}^{t}\}_{k,o},
        \{G_o^{t}\}_{o}
    \right).
    \label{eq:search_state}
\end{equation}

A generated expression first undergoes deterministic checks for
grammar, syntax, duplicate expressions, constraints, and
executability. Invalid candidates do not enter the evolutionary
DAG or consume the valid-evaluation budget. A valid candidate is
evaluated on the training and validation periods, producing an
event $e_{t+1}$ that records its performance, operator, parent
relations, expression change, and generalization evidence. The
search state then evolves according to
\begin{equation}
    \Omega^{t+1}
    =
    \mathcal{U}
    \left(
        \Omega^{t},
        f_{t+1},
        e_{t+1}
    \right),
    \label{eq:state_update}
\end{equation}
where $\mathcal{U}$ updates the lineage DAG, active population,
and elite archive, and invokes the primary-lineage and global
prior updates in Equation~\ref{eq:prior_update} when the
deterministic trigger is satisfied.

The updated lineage state carries operator experience across
generations while determining how subsequent search opportunities
are allocated. It therefore connects persistent LLM guidance with
lineage-level population control within a single evolutionary
loop.
\section{Experiments}

We evaluate \textsc{LineageEvo} on formulaic alpha factor mining to answer the following questions: (1) whether lineage-aware search finds more predictive and stable factors; (2) whether lineage and global priors provide useful guidance for LLM-based variation; (3) whether Lineage-Stratified Parent Selection (LSPS) improves factor diversity and reduces search concentration; and (4) whether \textsc{LineageEvo} generates higher-quality factors efficiently in the search.

\subsection{Experimental Setup}

\paragraph{Evaluation metrics.}
We report IC and ICIR to measure factor predictiveness, where IC denotes the average cross-sectional correlation between factor values and future returns, and ICIR reflects the temporal stability of IC. We also report portfolio-level metrics, including annualized return (AR), information ratio (IR), and maximum drawdown (MDD). AR and IR measure profitability and risk-adjusted performance, while MDD captures downside risk. Detailed metric definitions are provided in the supplementary material.

\paragraph{Dataset.}
We conduct experiments on two market universes: CSI 500 in the Chinese A-share market and S\&P 500 in the U.S. stock market. Both markets use the same fixed temporal split: data from 2015-01-01 to 2020-12-31 for training, 2021-01-01 to 2022-04-30 for validation, and 2022-05-01 to 2026-04-30 for testing.

\begin{table*}[ht]
\centering
\footnotesize
\setlength{\tabcolsep}{3.2pt}
\renewcommand{\arraystretch}{1.16}
\begin{tabular}{
    @{\hspace{4pt}}
    l
    ccccc
    @{\hspace{10pt}}
    ccccc
    @{\hspace{4pt}}
}
\toprule
&
\multicolumn{5}{c}{\textbf{CSI 500}}
&
\multicolumn{5}{c}{\textbf{S\&P 500}} \\
\cmidrule(l{1pt}r{1pt}){2-6}
\cmidrule(l{1pt}r{1pt}){7-11}
Method
& IC $\uparrow$
& ICIR $\uparrow$
& AR $\uparrow$
& IR $\uparrow$
& MDD $\downarrow$
& IC $\uparrow$
& ICIR $\uparrow$
& AR $\uparrow$
& IR $\uparrow$
& MDD $\downarrow$ \\
\midrule

Alpha158
& 0.0076 & 0.0787 & \underline{15.50\%}
& 0.6866 & 31.62\%
& 0.0081 & 0.0426 & 17.58\%
& 0.7715 & 21.03\% \\

GP
& 0.0074 & 0.0774 & 15.28\%
& 0.6861 & 33.08\%
& 0.0055 & 0.0575 & 21.38\%
& \underline{1.0958} & 20.55\% \\

AlphaGen
& 0.0345 & 0.2215 & 15.04\%
& 0.7176 & 36.31\%
& 0.0099 & 0.0587 & 20.38\%
& 0.9709 & 23.87\% \\

AlphaSAGE
& 0.0230 & 0.1855 & 13.71\%
& \underline{0.7654} & 21.31\%
& 0.0118 & \underline{0.0668} & 18.32\%
& 0.8539 & 21.62\% \\

RD-Agent
& 0.0395 & 0.2253 & 14.14\%
& 0.7199 & 29.48\%
& 0.0094 & 0.0464 & 11.05\%
& 0.7642 & 20.61\% \\

AlphaPROBE
& 0.0385 & 0.2298 & 11.87\%
& 0.7587 & \underline{20.48\%}
& \underline{0.0119} & 0.0622 & 5.86\%
& 0.4597 & 25.55\% \\

EoH
& \underline{0.0399} & \underline{0.2493} & 9.51\%
& 0.5715 & 21.50\%
& 0.0072 & 0.0404 & \underline{23.13\%}
& 0.9682 & 21.83\% \\

ReEvo
& 0.0303 & 0.1950 & 13.66\%
& 0.6030 & 42.98\%
& 0.0047 & 0.0267 & 16.10\%
& 0.7397 & \underline{18.94\%} \\

\textbf{LineageEvo}
& \textbf{0.0512} & \textbf{0.2740} & \textbf{15.55\%}
& \textbf{0.9856} & \textbf{19.39\%}
& \textbf{0.0134} & \textbf{0.0907} & \textbf{27.00\%}
& \textbf{1.1693} & \textbf{17.05\%} \\

\bottomrule
\end{tabular}
\caption{
Main results on CSI 500 and S\&P 500. Bold and underlined values
denote the best and second-best results.
}
\label{tab:main_results}
\end{table*}

\paragraph{Baselines.}
We compare \textsc{LineageEvo} with representative baselines from three categories:
(1) \textbf{traditional factor mining methods}, including Alpha158 \citep{qlib}, GP \citep{koza1992gp}, AlphaGen \citep{alphagen}, and AlphaSAGE \citep{alphasage};
(2) \textbf{LLM-based factor mining methods}, including RD-Agent \citep{rdagent} and AlphaPROBE \citep{alphaprobe}; and
(3) \textbf{LLM-EA methods}, including EoH \citep{eoh} and ReEvo \citep{reevo}.
Detailed descriptions of the baselines are provided in the supplementary material.

\paragraph{Implementation details.}
For all LLM-based methods, we use
\texttt{deepseek-v4-flash}. \textsc{LineageEvo} initializes each
run with 10 LLM-generated seed roots and targets three mutation
and two crossover offspring per generation until reaching
$B=200$ valid evaluated factors. \textsc{LineageEvo} uses an active pool of 50
factors and an elite archive of 20, with
$(\alpha,\beta,\gamma)=(0.2,0.8,0.1)$, and
$\lambda_{\mathrm{LSPS}}=0.8$. We use IC as the optimization metric $\mathcal{M}$ in the
reported experiments. The prior improvement/degradation thresholds are
$(0.003,0.30)/(0.012,0.60)$ on CSI 500 and
$(0.0015,0.20)/(0.006,0.40)$ on S\&P 500, where each pair gives
the absolute and relative thresholds. All methods use the same Qlib
TopK-Dropout backtest with top-$K=50$, $n_{\mathrm{drop}}=5$,
daily rebalancing, and matched risk and transaction-cost
settings \citep{qlib}. Complete prompt templates, DSL definitions,
prior schemas, and validation rules are provided in the
supplementary material.

\subsection{Main Results}

Table~\ref{tab:main_results} compares \textsc{LineageEvo} with
representative traditional, LLM-based, and LLM-EA factor mining
methods on both factor-level and portfolio-level metrics. For each method, we select the top 20 factors from its final
factor pool according to the
method-specific selection criterion when applicable, combine
them into a factor set, and evaluate the resulting combined
signal under the same evaluation pipeline.

\textsc{LineageEvo} achieves the best result on all five metrics
in both markets. On CSI 500, it obtains an IC of 0.0512 and an
ICIR of 0.2740, together with the highest AR (15.55\%) and IR
(0.9856) and the lowest MDD (19.39\%). On S\&P 500, it similarly
achieves the highest IC (0.0134), ICIR (0.0907), AR (27.00\%),
and IR (1.1693), while reducing MDD to 17.05\%. The consistent
improvements in IC and ICIR indicate that the discovered factors
provide both stronger out-of-sample predictiveness and greater
temporal stability. Meanwhile, the gains in AR and IR together
with lower drawdown show that these predictive improvements
can translate into more profitable and risk-controlled portfolios
across both markets. Overall, the results support the effectiveness
of combining lineage-conditioned variation with lineage-level
search allocation.

\begin{figure}[ht]
\centering
\includegraphics[width=0.85\columnwidth, 
trim=8 8 10 10,
    clip]{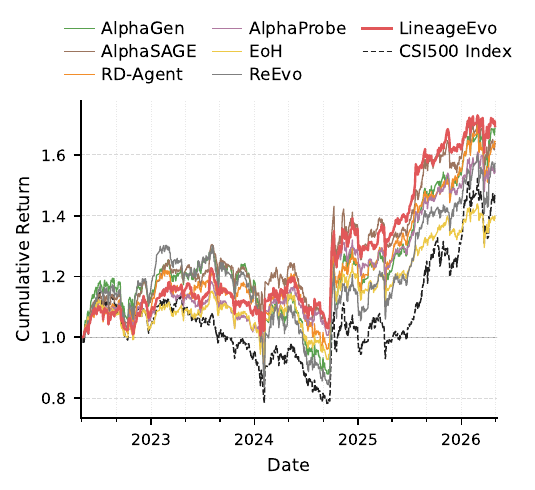}
\caption{Cumulative return curves of different factor mining
methods on CSI 500 during the test period.}
\label{fig:csi500_backtest}
\end{figure}

Figure~\ref{fig:csi500_backtest} shows that while achieving high predictive power, \textsc{LineageEvo}
maintains competitive cumulative returns throughout the test
period and becomes one of the strongest-performing methods in the
later stage, ultimately finishing with the highest cumulative
return among the compared approaches.

\subsection{Ablation on Search Priors}

We isolate the contribution of each component in the prior
system. \textsc{LineageEvo} uses lineage priors to capture
lineage-scoped experience and global priors to provide
cross-lineage regularization. Since mutation and crossover
require different forms of evolutionary experience, we also
examine the contributions of their corresponding
operator-specific priors.

We consider the following variants:
\begin{itemize}
\item \textbf{Full}: the complete \textsc{LineageEvo} framework.
\item \textbf{w/o all priors}: removes both lineage and global
priors.
\item \textbf{w/o lineage priors}: removes lineage mutation and
crossover priors while retaining global priors.
\item \textbf{w/o global priors}: removes global mutation and
crossover priors while retaining lineage priors.
\item \textbf{w/o mutation priors}: removes both lineage and
global mutation priors while retaining crossover priors.
\item \textbf{w/o crossover priors}: removes both lineage and
global crossover priors while retaining mutation priors.
\end{itemize}

\begin{table}[ht]
\centering
\renewcommand{\arraystretch}{1.08}
\begin{tabular}{@{}l@{\hspace{18pt}}c@{\hspace{18pt}}c@{}}
\toprule
Variant
& IC $\uparrow$
& ICIR $\uparrow$ \\
\midrule
w/o all priors
& 0.0384
& 0.1994 \\

w/o lineage priors
& 0.0390
& 0.2041 \\

w/o global priors
& \underline{0.0495}
& 0.2579 \\

w/o mutation priors
& 0.0438
& \underline{0.2673} \\

w/o crossover priors
& 0.0445
& 0.2362 \\

Full \textsc{LineageEvo}
& \textbf{0.0512}
& \textbf{0.2740} \\
\bottomrule
\end{tabular}
\caption{Ablation study of the search-prior components on CSI 500.}
\label{tab:prior_ablation}
\end{table}

The ablation results confirm the effectiveness of the prior
system. Removing all priors reduces IC from 0.0512 to 0.0384 and
ICIR from 0.2740 to 0.1994, showing that prior-guided variation
substantially improves factor quality. Removing lineage priors produces a similarly large
degradation, indicating that lineage-scoped experience is the
main source of useful guidance, while global priors alone
cannot fully capture the transformation patterns specific to
each lineage. The operator-specific
ablations further reveal the important roles of mutation and
crossover priors.

\subsection{Factor Diversity and Search Concentration}

We evaluate how LSPS affects predictive performance, lineage
concentration, and expression diversity in the active factor
pool. We compare \textsc{LineageEvo} with a variant that removes
LSPS and samples parents directly from the active pool without
lineage stratification. Lineage entropy measures how evenly
factors are distributed across lineages, average pairwise AST
distance measures expression-level structural diversity, and the
effective number of structures estimates the number of
non-redundant structural families in the pool.

\begin{table}[ht]
\centering
\footnotesize
\setlength{\tabcolsep}{2.4pt}
\renewcommand{\arraystretch}{1.08}
\begin{tabular}{@{}lccccc@{}}
\toprule
Method
& IC $\uparrow$
& ICIR $\uparrow$
& \shortstack{Lineage\\Entropy $\uparrow$}
& \shortstack{AST\\Dist. $\uparrow$}
& \shortstack{Eff.\\Struct. $\uparrow$} \\
\midrule
w/o LSPS
& 0.0381
& 0.2415
& 0.405
& 0.509
& 37.89 \\

\textsc{LineageEvo}
& \textbf{0.0512}
& \textbf{0.2740}
& \textbf{2.154}
& \textbf{0.747}
& \textbf{41.90} \\
\bottomrule
\end{tabular}
\caption{Effect of LSPS on predictive performance and diversity on
CSI 500. Higher values are better.}
\label{tab:factor_diversity}
\end{table}

As shown in Table~\ref{tab:factor_diversity}, LSPS improves IC
from 0.0381 to 0.0512 and ICIR from 0.2415 to 0.2740, while
substantially increasing all three diversity measures. This
joint improvement reflects the quality--exploration trade-off
built into LSPS. The quality component continues to allocate
search opportunities to lineages containing strong factors,
allowing promising structures to be refined across successive
generations. Meanwhile, the lineage-balance component prevents
these branches from monopolizing the search and preserves
opportunities for underrepresented lineages that may contain
distinct but still promising directions. Consequently, the final
factor pool retains high-performing descendants while covering a
broader and less redundant set of structures. Such a pool is more
likely to provide complementary candidates for factor
combination, which is consistent with the higher IC and ICIR
achieved by the full method
\citep{sudholt2018diversity,gabor2018inheritance,
alphagen,alphaforge}.

\subsection{Search Efficiency}

We analyze whether \textsc{LineageEvo} can discover high-quality factors more efficiently under the same budget. For each method, we track the top-5 mean-so-far absolute IC during the search process. Specifically, at each evaluation checkpoint, we collect all valid factors generated so far, select the five factors with the highest absolute IC, and report their average IC.

\begin{figure}[h]
\centering
\includegraphics[width=0.95\columnwidth,
trim=0 10 0 6,
    clip]{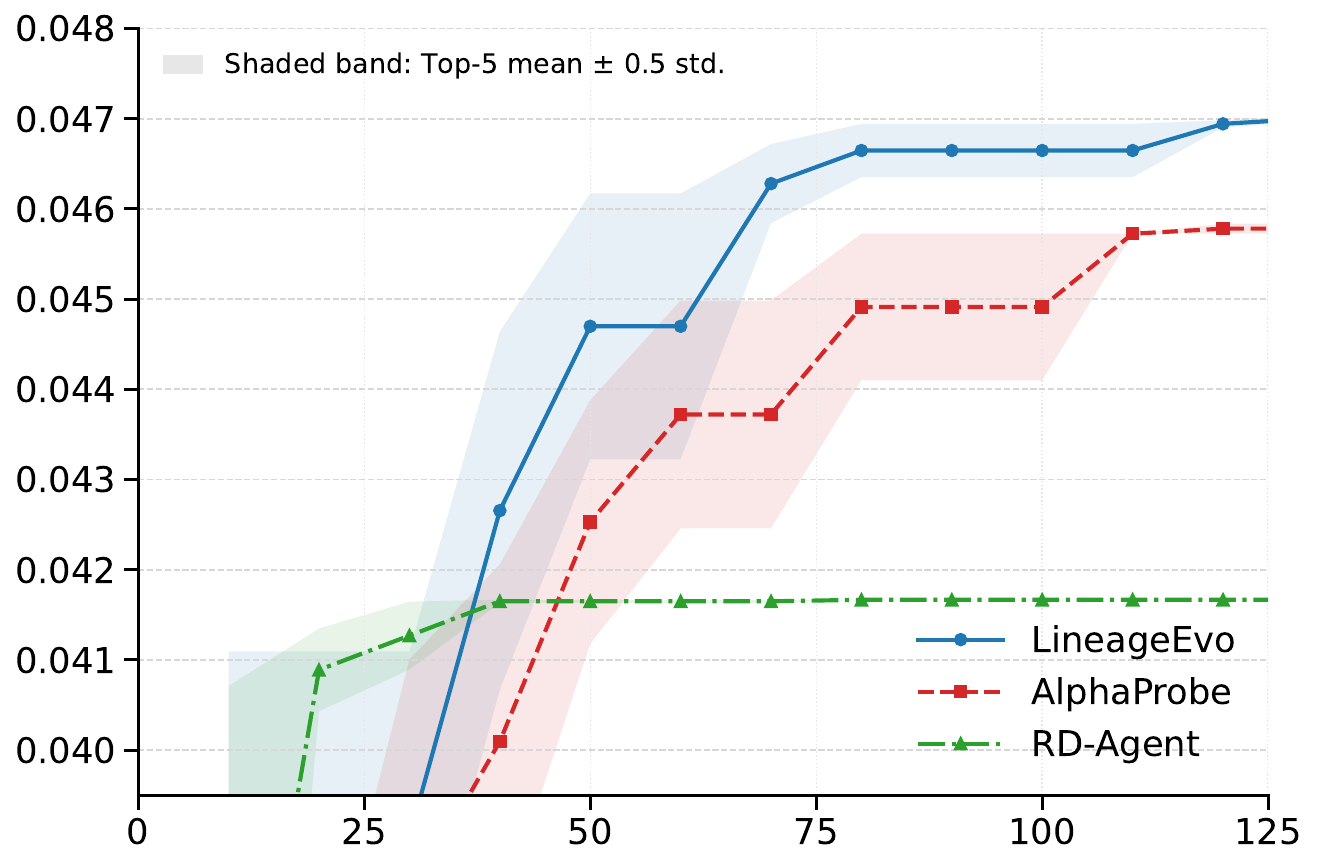}
\caption{Search efficiency comparison on CSI 500. The x-axis denotes the number of valid evaluated factors, and the y-axis reports the top-5 mean-so-far absolute IC.}
\label{fig:search_efficiency}
\end{figure}

Figure~\ref{fig:search_efficiency} shows different search behaviors among the three methods. RD-Agent improves quickly in the early stage but soon saturates at a relatively low top-5 mean-so-far IC, suggesting that it can rapidly identify some usable factors but lacks a mechanism to continuously exploit promising search directions. AlphaPROBE improves more gradually and eventually reaches a stronger factor set than RD-Agent, but its improvement requires more evaluations. In contrast, \textsc{LineageEvo} maintains a consistently higher top-5 mean-so-far IC throughout most of the search process and reaches a strong plateau within about 75 valid evaluations. This pattern is consistent with the intended role of
lineage-conditioned priors in preserving useful search
guidance across generations.

\section{Conclusion}

In this paper, we proposed \textsc{LineageEvo}, an LLM-EA
framework that uses lineage as an intermediate organizational
unit for evolutionary factor mining. \textsc{LineageEvo}
combines lineage-conditioned priors for guiding LLM variation with Lineage-Stratified Parent Selection for
distributing search opportunities across lineages.
The resulting framework discovers highly predictive and stable
factors, achieves strong investment returns, and preserves a more
diverse factor pool. Future work may compare lineage with other
intermediate organizational units, such as 
clusters, niches, and retrieval neighborhoods, to study their
respective advantages in experience reuse and search control.

\bibliography{refs}

\appendix
\setcounter{secnumdepth}{1}

\section{Case Study}
\label{app:case_study}

\begin{figure*}[t]
    \centering
    \includegraphics[width=0.92\textwidth]
    {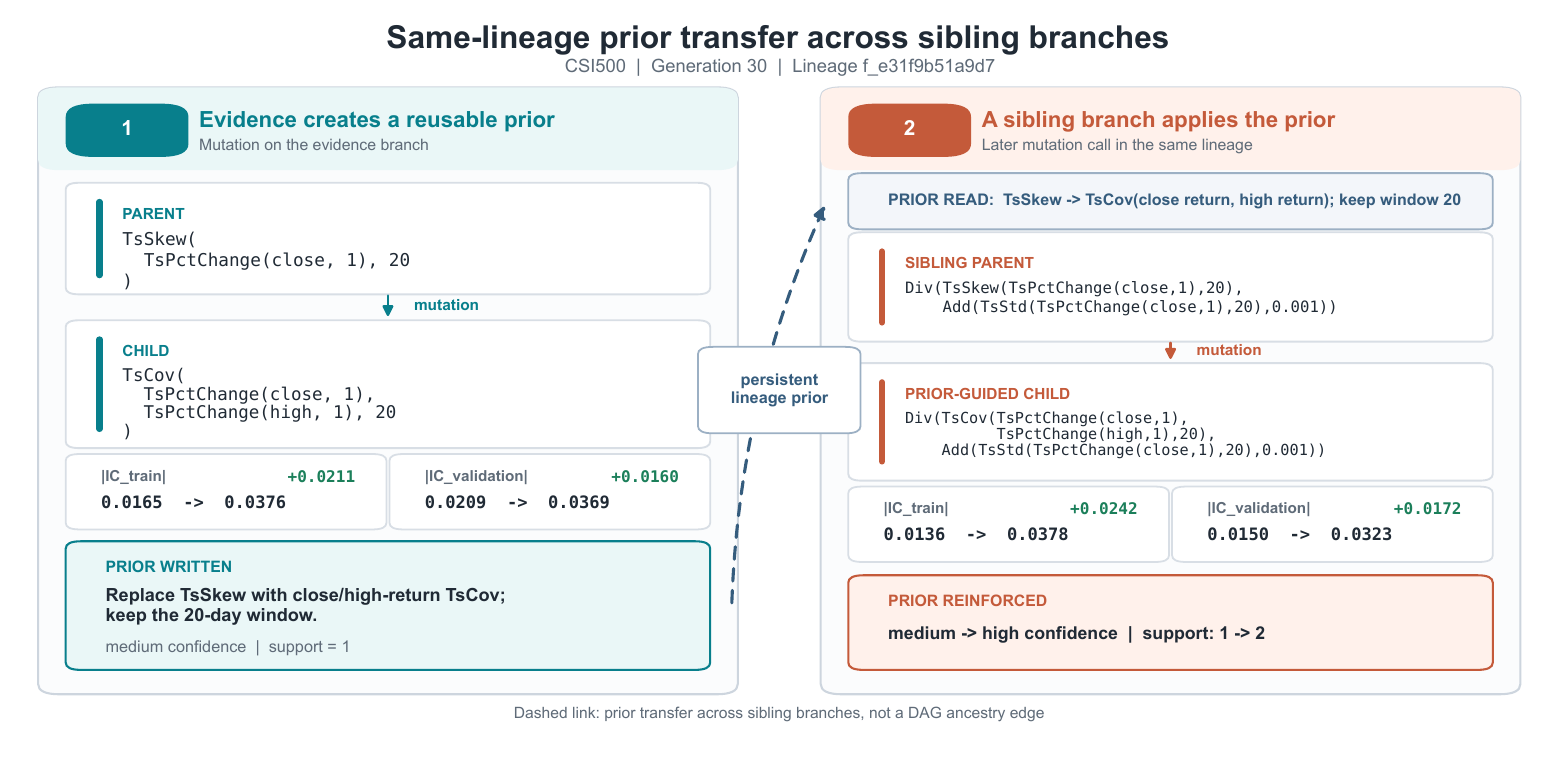}
    \caption{A same-lineage prior-reuse case. Event 1 replaces
    \texttt{TsSkew} with a close/high-return \texttt{TsCov},
    triggering a structured prior rewrite. Event 2 receives this
    prior on a sibling branch, applies the same replacement while
    preserving its denominator, and strengthens the pattern from
    medium to high confidence. The dashed connector denotes prior
    transfer.}
    \label{fig:case_evolution}
\end{figure*}

\paragraph{Illustrative trace.}
Figure~\ref{fig:case_evolution} shows two sequential mutation
events on sibling branches of the same lineage in a CSI500 run.
At generation 30, Event 1 replaces \texttt{TsSkew} with
\texttt{TsCov} between close and high returns, increasing
validation $|\mathrm{IC}|$ from 0.0209 to 0.0369. This event
triggers both lineage and global prior rewriting. Event 2 then
receives the updated priors before generating a related mutation
on another branch of the same lineage. All reported values are
absolute IC strengths.

\paragraph{Rendered prior guidance.}
Before Event 2, Event 1 had already triggered lineage and global
prior rewriting. The following excerpt shows the actual
prior-related guidance rendered into the mutation prompt.

\begin{promptbox}[label={prompt:case-prior}]{Actual Prior Guidance Before Event 2}
Mutation experience for this lineage:
- Effective pattern:
  Replacing TsSkew with TsCov between close returns and a second
  price series, while preserving the 20-day rolling window,
  produced a strong validation IC improvement.

- Failed patterns:
  Avoid replacing TsSkew with TsCorr or TsIr.
  Avoid introducing raw volume terms or EMA smoothing.
  Preserve the close-return distributional backbone.

- Hint:
  TsCov with a second price-return series may be useful, while
  correlation, smoothing, and volume-based modifications have
  repeatedly degraded this lineage.

- Bias warning: medium


Global mutation experience:
- Percentage-change inputs have repeatedly strengthened covariance
  and correlation structures across lineages.

- TsCov/TsCorr structures should operate on transformed return-like
  inputs rather than raw price or volume levels.

- Avoid TsIr substitutions, unnecessary smoothing, and volume-delta
  multipliers.

- Last updated generation: 30


Prior fusion decision:
- Lineage prior weight: 0.45
- Global prior weight: 0.55
- Balance lineage-specific evidence with global operator guidance.
\end{promptbox}

The example illustrates the complementary roles of the two prior
levels. The lineage prior provides factor-family-specific evidence
about successful and failed transformations, whereas the global
prior contributes patterns observed across lineages. Their
relative emphasis is determined by the fusion policy before the
guidance is inserted into the candidate-generation prompt.

\section{Metric Definitions}
\label{app:metrics}

Let $f$ denote a formulaic factor, $z^f_{i,t}$ its value for stock $i$ at date $t$, and $r_{i,t+\Delta}$ the future return used as the prediction target. Let $\mathcal{U}_t$ be the tradable stock universe at date $t$. The factor-level and portfolio-level metrics follow standard formulaic-alpha and Qlib evaluation practice \citep{formulaic101,qlib}.

\paragraph{Information Coefficient (IC).}
We compute the daily cross-sectional Spearman correlation between factor values and future returns:
\begin{equation}
    \mathrm{IC}_t(f)
    =
    \mathrm{corr}^{\mathrm{Spearman}}_{i\in\mathcal{U}_t}
    \left(z^f_{i,t},r_{i,t+\Delta}\right).
\end{equation}
The reported IC is the temporal mean of daily cross-sectional IC:
\begin{equation}
    \mathrm{IC}(f)
    =
    \frac{1}{|\mathcal{T}|}
    \sum_{t\in\mathcal{T}}\mathrm{IC}_t(f).
\end{equation}
Under this implementation, IC is equivalent to RankIC.

\paragraph{Information Coefficient Information Ratio (ICIR).}
ICIR measures the temporal stability of daily IC:
\begin{equation}
    \mathrm{ICIR}(f)
    =
    \frac{\operatorname{mean}_{t\in\mathcal{T}}[\mathrm{IC}_t(f)]}
    {\operatorname{std}_{t\in\mathcal{T}}[\mathrm{IC}_t(f)]+\epsilon},
\end{equation}
where $\epsilon$ is used only to avoid division by zero.

\paragraph{Annualized Return (AR).}
Let $R_t$ denote the daily return series passed to Qlib's risk-analysis routine and let $N$ denote its daily annualization scaler. Following Qlib's default arithmetic accumulation, we report
\begin{equation}
    \mathrm{AR}
    =
    N\,\operatorname{mean}_{t\in\mathcal{T}}[R_t].
\end{equation}

\paragraph{Information Ratio (IR).}
Risk-adjusted return is measured as
\begin{equation}
    \mathrm{IR}
    =
    \frac{\operatorname{mean}_{t\in\mathcal{T}}[R_t]}
    {\operatorname{std}_{t\in\mathcal{T}}[R_t]+\epsilon}
    \sqrt{N}.
\end{equation}

\paragraph{Maximum Drawdown (MDD).}
Under arithmetic return accumulation, let $C_t=\sum_{\tau\leq t}R_\tau$. Qlib's signed drawdown statistic is
\begin{equation}
    \mathrm{MDD}_{\mathrm{signed}}
    =
    \min_{t\in\mathcal{T}}
    \left(C_t-\max_{\tau\leq t}C_\tau\right).
\end{equation}
When a table reports MDD with a downward arrow and nonnegative entries, we use the magnitude $|\mathrm{MDD}_{\mathrm{signed}}|$; when signed values are shown, values closer to zero are better.

\paragraph{Validation-Oriented Test Metrics.}
The factor direction is fixed using validation information before test evaluation. For a metric $\mathcal{M}$, the oriented test IC is
\begin{equation}
    \mathrm{IC}^{\mathrm{oriented}}_{\mathrm{test}}(f)
    =
    \operatorname{sign}\!\left(\mathcal{M}_{\mathrm{val}}(f)\right)
    \mathrm{IC}_{\mathrm{test}}(f).
\end{equation}
No test statistic is used for factor orientation, parent selection, prior updating, or final factor selection.

\paragraph{Top-5 Mean-so-Far IC.}
Let $\mathcal{F}_k$ denote the valid factors discovered within the first $k$ valid evaluations. Search efficiency is summarized using validation IC:
\begin{equation}
    \mathrm{Top5IC}_{\mathrm{val}}(k)
    =
    \frac{1}{5}
    \sum_{f\in\mathrm{Top5}(\mathcal{F}_k)}
    \left|\mathrm{IC}_{\mathrm{val}}(f)\right|,
\end{equation}
where $\mathrm{Top5}(\mathcal{F}_k)$ contains the five factors with the largest absolute validation IC discovered so far. Invalid and duplicate attempts do not advance the horizontal-axis evaluation count.

\paragraph{Lineage Entropy.}
Let $p_L$ be the fraction of factors in the evaluated population state that belong to lineage $L$. We compute
\begin{equation}
    H_{\mathrm{lineage}}
    =
    -\sum_L p_L\log p_L.
\end{equation}
Higher entropy indicates a more balanced allocation across lineages \citep{sudholt2018diversity,gabor2018inheritance}.

\paragraph{Average Pairwise AST Distance.}
We measure expression-level diversity with normalized tree-edit distance between factor abstract syntax trees (ASTs). For a factor set $\mathcal{S}$,
\begin{equation}
    D_{\mathrm{AST}}(\mathcal{S})
    =
    \frac{2}{|\mathcal{S}|(|\mathcal{S}|-1)}
    \sum_{i<j}d_{\mathrm{AST}}(f_i,f_j).
\end{equation}
A larger value indicates greater structural dissimilarity.

\paragraph{Effective Number of Structures.}
For each factor $f$, let $\sigma(f)$ denote the complete
structural signature of its normalized AST. We apply the same
canonical normalization used for expression deduplication and
retain all nodes and ordered parent--child relations in the
resulting tree. Two factors belong to the same structural group
if and only if
\begin{equation}
    \sigma(f_i)=\sigma(f_j).
\end{equation}
Thus, the groups correspond to exact normalized AST structures
rather than distance-based clusters or partial motifs. Let $p_c$
be the empirical frequency of structural signature $c$ in the
factor set. We report
\begin{equation}
    N_{\mathrm{eff}}
    =
    \exp\!\left(-\sum_c p_c\log p_c\right),
\end{equation}
which gives the number of equally represented structures with
the same structural entropy.

\section{Baseline Descriptions}
\label{app:baselines}

We compare \textsc{LineageEvo} with representative baselines from traditional factor mining, LLM-based factor mining, and LLM-EA factor mining.

\paragraph{Alpha158.}
Alpha158 is a standard hand-crafted factor set provided in Qlib and serves as a non-search reference \citep{qlib}.

\paragraph{Genetic Programming (GP).}
GP provides a classical non-LLM evolutionary baseline based on symbolic mutation, crossover, and fitness selection \citep{koza1992gp}.

\paragraph{AlphaGen.}
AlphaGen generates synergistic formulaic alpha collections through reinforcement learning \citep{alphagen}.

\paragraph{AlphaSAGE.}
AlphaSAGE performs structure-aware factor mining with GFlowNet-based exploration \citep{alphasage}.

\paragraph{RD-Agent.}
R\&D-Agent-Quant is an LLM-based multi-agent framework for data-centric factor and model optimization \citep{rdagent}.

\paragraph{AlphaPROBE.}
AlphaPROBE combines principled retrieval with on-graph biased evolution for alpha mining \citep{alphaprobe}.

\paragraph{EoH.}
Evolution of Heuristics uses LLMs to iteratively generate, evaluate, and evolve heuristic programs \citep{eoh}.

\paragraph{ReEvo.}
ReEvo uses reflective evolution to guide LLM-based heuristic search \citep{reevo}.

\begin{table*}[t]
\centering
\footnotesize
\setlength{\tabcolsep}{4pt}
\begin{tabular}{p{0.14\textwidth}p{0.28\textwidth}p{0.50\textwidth}}
\toprule
Component & Parameter & Setting \\
\midrule
LLM & Backend & \texttt{deepseek-v4-flash} through an OpenAI-compatible client \\
LLM & Temperature / maximum tokens & $0.7$ / $8000$ \\
LLM & Timeout / retries / retry wait & $120$ s / $5$ / $5$ s \\
Search & Valid-evaluation budget & $B=200$ \\
Search & Initial lineage roots & $10$ LLM-generated seed factors \\
Search & Valid mutation / crossover slots per generation & $3$ / $2$ \\
Search & Maximum attempts per slot & $5$ \\
Population & Population capacity / elite archive capacity & $50$ / $20$ \\
Population & Maximum share of one lineage & $0.40$ \\
Expression & Prompt / deterministic length limits & $40$ / $50$ \\
Robust score & $(\alpha,\beta,\gamma)$ & $(0.2,0.8,0.1)$ \\
LSPS & Quality / diversity mixture weights & $(0.8,0.2)$ \\
Backtest & Top-$K$ / $n_{\mathrm{drop}}$ / risk degree
& $50$ / $5$ / $0.95$ \\

Backtest & Rebalancing frequency
& Daily \\

Backtest & Buy / sell cost
& $\texttt{open\_cost}=0.0015$ /
  $\texttt{close\_cost}=0.0025$ \\

Backtest & Minimum transaction cost
& $\texttt{min\_cost}=5$ \\

Finalization & Number of selected factors
& $1$, $5$, or $20$, as specified \\
\bottomrule
\end{tabular}
\caption{Main implementation and evaluation parameters.}
\label{tab:implementation_parameters}
\end{table*}

\section{Experimental Protocol and Implementation Details}
\label{app:implementation}

\paragraph{Markets, target, and temporal splits.}
We evaluate on CSI500 and S\&P 500 using daily data and a common Qlib-compatible pipeline \citep{qlib}. The prediction target is the next-period return implemented as
\begin{equation}
    \texttt{Ref(\$close,-2)/Ref(\$close,-1)-1}.
\end{equation}
For both markets, 2015-01-01 to 2020-12-31 is used for training, 2021-01-01 to 2022-04-30 for validation, and 2022-05-01 to 2026-04-30 for testing. Training and validation data may be used during search; the test period is accessed only after the final selection rule and factor set are fixed.

\paragraph{Core experimental parameters.}
Table~\ref{tab:implementation_parameters} summarizes the shared configuration. Values associated with final Top-$K$ selection vary only when the corresponding Top-1, Top-5, or Top-20 result is explicitly reported.

\paragraph{Candidate generation schedule.}
Each LLM call returns exactly one candidate expression. A generation targets three valid mutation offspring and two valid crossover offspring. Each slot permits at most five generation attempts. Invalid, duplicate, and failed-evaluation candidates are logged but do not count toward the valid-evaluation budget.

\paragraph{Expression grammar and deterministic validation.}
Candidates may use only the configured market variables, constants, and DSL operators. Validation checks JSON/expression parsing, permitted variables and operators, constants and window arguments, normalized-expression duplication, expression length, Qlib executability, empty or constant outputs, and NaN/Inf values. Only valid evaluated candidates are inserted into the lineage DAG.

\paragraph{Optimization metric and robust quality score.}
The configured optimization metric $\mathcal{M}$ is IC or ICIR.
For factors satisfying
$|\mathcal{M}_{\mathrm{tr}}(f)|
+|\mathcal{M}_{\mathrm{val}}(f)|>0$, let
\begin{equation}
    T(f)=|\mathcal{M}_{\mathrm{tr}}(f)|,
    \qquad
    V(f)=|\mathcal{M}_{\mathrm{val}}(f)|.
\end{equation}
Parent, lineage-quality, and final-selection decisions use
\begin{equation}
    q(f)
    =
    0.2T(f)+0.8V(f)
    -0.1
    \left|
        \mathcal{M}_{\mathrm{tr}}(f)
        -
        \mathcal{M}_{\mathrm{val}}(f)
    \right|.
\end{equation}
The validation term receives the largest weight, while the signed
discrepancy penalizes both strength disagreement and directional
reversals between training and validation.

\paragraph{Lineage-stratified sampling.}
For lineage $k$, let $b_k=\max_{f\in\mathcal{P}_k}q(f)$ and $d_k=(|\mathcal{P}_k|+1)^{-1}$. The lineage distribution is
\begin{equation}
    \pi_{\mathrm{lin}}(k)
    =
    0.8\frac{b_k}{\sum_j b_j}
    +
    0.2\frac{d_k}{\sum_j d_j}.
\end{equation}
After sampling lineage $k$, a parent is sampled according to
\begin{equation}
    \pi_{\mathrm{par}}(f\mid k)
    =
    \frac{q(f)}{\sum_{g\in\mathcal{P}_k}q(g)}.
\end{equation}
The population updater additionally limits the fraction of active factors belonging to any one lineage to $0.40$.

\paragraph{Mutation, crossover, and lineage inheritance.}
Mutation selects one lineage and one parent. Crossover selects an anchor lineage and parent, then preferentially samples the partner from another lineage using the same quality--diversity policy over the remaining lineages. The parent sampled from the first
selected lineage becomes the primary parent. The child inherits the primary parent's lineage, while the other parent is retained through a secondary crossover edge. Operationally, crossover replaces or enriches a subtree of the primary parent with a subtree-level structure from the secondary parent.

\paragraph{Operator-specific priors and fusion.}
Each lineage maintains separate mutation and crossover priors, and the search maintains corresponding global operator priors. Structured prior states are rendered into concise natural-language guidance before candidate generation. A deterministic fusion policy determines the relative emphasis on lineage-specific and global experience; the resulting context is soft guidance rather than a hard expression constraint. All compared variants share the same fusion and prompt infrastructure unless a prior source is explicitly removed.

\paragraph{Event-triggered prior rewriting.}
Prior rewriting is considered only after a valid candidate has
been evaluated. We take IC as an example. Let $\mathrm{IC}_{\mathrm{par}}$ and
$\mathrm{IC}_{\mathrm{child}}$ denote the IC values of the
reference parent and its offspring, respectively. For crossover,
the primary parent serves as the reference parent. All comparisons
use absolute IC strength, consistent with the search objective.

On CSI500, an improvement trigger is activated when
\begin{equation}
    |\mathrm{IC}_{\mathrm{child}}|
    -
    |\mathrm{IC}_{\mathrm{par}}|
    \geq
    \max\left(
        0.003,\,
        0.30|\mathrm{IC}_{\mathrm{par}}|
    \right),
\end{equation}
whereas a degradation trigger is activated when
\begin{equation}
    |\mathrm{IC}_{\mathrm{par}}|
    -
    |\mathrm{IC}_{\mathrm{child}}|
    \geq
    \max\left(
        0.012,\,
        0.60|\mathrm{IC}_{\mathrm{par}}|
    \right).
\end{equation}
S\&P 500 uses more sensitive thresholds. Its improvement and
degradation thresholds are
$\max(0.0015,0.20|\mathrm{IC}_{\mathrm{par}}|)$ and
$\max(0.006,0.40|\mathrm{IC}_{\mathrm{par}}|)$, respectively.

When either trigger is activated, the rewriter receives the
previous structured prior together with a structured summary of
the evolutionary event and returns a complete updated state rather
than a partial patch. A deterministic manager checks the output
schema, merges redundant entries, removes low-value content, and
retains the previous prior if rewriting fails. If neither trigger
is activated, the prior remains unchanged. This event-triggered
mechanism differs from unrestricted free-form reflection
\citep{reevo}.

\paragraph{Population and elite archive.}
The population is used for future search allocation and is organized by lineage. The elite archive is updated independently and preserves the strongest evaluated factors.

\paragraph{Final factor selection and orientation.}
Final factors are selected primarily from the elite archive using the same robust score $q(f)$ defined above. Expressions are deduplicated after normalization. The sign of each selected factor is fixed using the corresponding validation metric, and no test statistic participates in selection or orientation. For Top-$K$ evaluation, the $K$ highest-scoring distinct factors are selected directly. When $K>1$, oriented factor values are normalized cross-sectionally each day and combined with equal weights.

\paragraph{Backtesting protocol.}
Portfolio evaluation uses Qlib's
\texttt{TopkDropoutStrategy} with
$\texttt{topk}=50$,
$\texttt{n\_drop}=5$,
$\texttt{risk\_degree}=0.95$, and daily execution
\citep{qlib}. The transaction-cost configuration uses the Qlib
default values:
$\texttt{open\_cost}=0.0015$ for purchases,
$\texttt{close\_cost}=0.0025$ for sales, and
$\texttt{min\_cost}=5$ per transaction. The same backtesting and cost
configuration is applied consistently across methods within each
market. AR, IR, and MDD are computed from the cost-adjusted return
series produced by the shared backtesting pipeline.

\paragraph{Ablation configurations.}
The prior ablations preserve the search budget, LLM backend, LSPS policy, validation protocol, and all non-ablated components. \emph{w/o all priors} removes lineage and global prior context; \emph{w/o lineage priors} retains only global operator priors; \emph{w/o global priors} retains only lineage-specific priors; \emph{w/o mutation priors} removes both lineage and global mutation guidance; and \emph{w/o crossover priors} analogously removes crossover guidance. The LSPS ablation replaces lineage-first allocation with score-weighted sampling over the complete population while leaving candidate generation and evaluation unchanged.

\section{Prompt Templates}
\label{app:prompts}

\textsc{LineageEvo} uses four principal prompt families for seed
generation, mutation, crossover, and prior rewriting. Each LLM
request contains one system message and one user message.
Candidate-generation prompts share the same factor DSL,
look-ahead prevention rules, expression-length constraints, and
JSON-format requirements. For readability, the templates below
retain the operator-specific instructions, runtime fields, and
output schemas while abbreviating repeated DSL definitions,
validation constraints, and long examples with bracketed
placeholders.

\begin{table*}[t]
\centering
\small
\setlength{\tabcolsep}{4pt}
\begin{tabular}{p{0.12\textwidth}p{0.18\textwidth}p{0.45\textwidth}p{0.17\textwidth}}
\toprule
Prompt & Trigger & Main runtime inputs & Required output \\
\midrule
Seed generation
& Population initialization
& Market, stock universe, factor DSL, expression-length target,
and existing or rejected seed expressions
& One factor with hypothesis and metadata in JSON \\

Mutation
& After selecting one parent
& Parent expression and metrics, lineage and global mutation
priors, search-control state, fusion decision, and retry feedback
& \texttt{\{"factor": "..."\}} \\

Crossover
& After selecting two parents
& Primary and secondary expressions and metrics, their lineage
priors, global crossover prior, fusion decision, and duplicate
feedback
& \texttt{\{"factor": "..."\}} \\

Prior rewriting
& After an informative valid evolutionary event
& Previous prior, parent and child evidence, expression
difference, update trigger, and target JSON schema
& A complete updated prior in JSON \\
\bottomrule
\end{tabular}
\caption{Summary of the principal prompts used by
\textsc{LineageEvo}.}
\label{tab:prompt_inventory}
\end{table*}

\paragraph{Shared candidate-generation instructions.}
Seed, mutation, and crossover prompts instruct the LLM to use
only the permitted market variables and DSL operators, avoid
future information, and generate exactly one executable factor.
The performance statistics provided to the LLM are absolute IC
or ICIR strengths; the sign used for final factor orientation is
not included in candidate-generation prompts.

\paragraph{Seed-generation prompt.}
The seed prompt encourages diversity across initial lineages and
asks for a financial hypothesis together with the expression.

\begin{promptbox}[label={prompt:seed}]{Seed-Generation Prompt (Compressed)}
SYSTEM

You are an expert quantitative researcher. Generate exactly one
 diverse and economically plausible seed factor using only the
 permitted DSL. Avoid look-ahead leakage, invalid expressions,
 and unnecessary complexity. Return strict JSON only.

USER

Market: <MARKET>
Stock universe: <UNIVERSE>
Allowed factor language:
[Shared DSL and operator definitions omitted.]

Maximum expression length: <LENGTH_TARGET>
Existing or previously rejected factors to avoid:
<EXISTING_OR_REJECTED_EXPRESSIONS>

Generate one seed based on a distinct financial hypothesis.
Return:
{
  "factor_name": "<NAME>",
  "hypothesis": "<FINANCIAL_HYPOTHESIS>",
  "factor": "<FACTOR_EXPRESSION>",
  "expected_signal_type": "<SIGNAL_TYPE>",
  "rationale": "<BRIEF_RATIONALE>",
  "risk_notes": "<RISKS>"
}
\end{promptbox}

\paragraph{Mutation prompt.}
The mutation prompt conditions the LLM on one parent, the
experience accumulated within its lineage, and the corresponding
global mutation prior. Program-computed control variables specify
whether the current search state calls for conservative or
broader variation.

\begin{promptbox}[label={prompt:mutation}]{Mutation Prompt (Compressed)}
SYSTEM

<CANDIDATE_SYSTEM_MESSAGE>
Generate exactly one mutation candidate and return strict JSON only.

USER

Parent factor: <PARENT_EXPRESSION>
Parent predictive strengths: <ABSOLUTE_PARENT_METRICS>

Lineage mutation prior:
<RENDERED_LINEAGE_MUTATION_PRIOR>

Global mutation prior:
<RENDERED_GLOBAL_MUTATION_PRIOR>

Search-control state:
quality_trend = <QUALITY_TREND>
stagnation_state = <STAGNATION_STATE>
mutation_strength = <MUTATION_STRENGTH>

Prior-fusion decision:
local_weight = <LOCAL_WEIGHT>
global_weight = <GLOBAL_WEIGHT>
instruction = <FUSION_INSTRUCTION>

Recent invalid reasons: <FAILURE_FEEDBACK>
Previous duplicate to avoid: <DUPLICATE_EXPRESSION>

Perform one coherent targeted mutation. Preserve useful parent
structure unless broader exploration is requested. Treat the
priors as soft guidance and do not override the program-computed
mutation strength.

[Shared DSL and validation constraints omitted.]

Return only:
{"factor": "<NEW_MUTATED_EXPRESSION>"}
\end{promptbox}

\paragraph{Crossover prompt.}
Crossover is implemented asymmetrically. The primary parent
provides the expression backbone, whereas the secondary parent
provides a local donor subtree. The prompt includes lineage
experience from both parents, but the fusion decision is centered
on the primary lineage.

\begin{promptbox}[label={prompt:crossover}]{Crossover Prompt (Compressed)}
SYSTEM

<CANDIDATE_SYSTEM_MESSAGE>
Generate exactly one crossover candidate and return strict JSON only.

USER

Primary parent: <PRIMARY_PARENT>
Primary predictive strengths: <PRIMARY_METRICS>
Secondary parent: <SECONDARY_PARENT>
Secondary predictive strengths: <SECONDARY_METRICS>

Primary-lineage crossover prior:
<PRIMARY_LINEAGE_PRIOR>

Secondary-lineage crossover prior:
<SECONDARY_LINEAGE_PRIOR>

Global crossover prior:
<GLOBAL_CROSSOVER_PRIOR>

Prior-fusion decision: <FUSION_DECISION>
Previous duplicate to avoid: <DUPLICATE_EXPRESSION>

Use the primary parent as the main expression tree. Import only a
proper non-root subtree or transformed structural motif from the
secondary parent. Do not copy the complete secondary expression.
Keep the imported material comparable in size and complexity to
the replaced primary subtree, and preserve the useful primary
backbone.

[Shared DSL and validation constraints omitted.]

Return only:
{"factor": "<NEW_CROSSOVER_EXPRESSION>"}
\end{promptbox}

\paragraph{Prior-rewriting prompt.}
Prior rewriting is triggered only after a valid candidate yields
informative evolutionary evidence. Rather than producing a
free-form reflection or a partial patch, the LLM rewrites the
complete structured prior, which is subsequently checked against
a deterministic schema.

\begin{promptbox}[label={prompt:rewrite}]{Prior-Rewriting Prompt (Compressed)}
SYSTEM

Rewrite a persistent structured prior, not a free-form reflection
or patch. Merge the previous knowledge with the new evidence,
preserve supported patterns, weaken contradicted patterns, merge
redundant entries, and keep the state compact. The reported
metrics are absolute predictive-strength values. Do not modify
deterministic validation, evaluation, or search control. Return
one complete JSON object matching the supplied schema, with no
Markdown.

USER

Target prior: <PRIOR_TARGET>
Previous structured prior:
<OLD_PRIOR_JSON>

New evolutionary evidence:
<EVALUATION_EVIDENCE_JSON>

Expression difference: <EXPRESSION_DIFF>
Update trigger: <TRIGGER>
Required JSON schema: <TARGET_JSON_SCHEMA>

Rewrite the complete prior according to the target-specific rules.
\end{promptbox}

The four rewriting targets use different schemas. A lineage
mutation prior records successful and failed mutation patterns,
a concise hint, and potential bias risks. A lineage crossover
prior records transferable and harmful patterns,
complementarity, heritable structures, and crossover risks.
Global mutation and crossover priors retain only patterns that
are considered transferable across lineages.

\begin{itemize}
\raggedright
\setlength{\itemsep}{2pt}

\item \textbf{Lineage mutation prior:}
\texttt{successful\_mutation\_patterns},
\texttt{failed\_mutation\_patterns},
\texttt{hint}, and \texttt{bias\_risk}.

\item \textbf{Lineage crossover prior:}
\texttt{transferable\_patterns},
\texttt{harmful\_patterns},
\texttt{complementarity\_profile},
\texttt{heritable\_structures},
\texttt{hint}, and \texttt{crossover\_risk}.

\item \textbf{Global mutation prior:}
\texttt{global\_successful\_mutation\_patterns},
\texttt{global\_failed\_mutation\_patterns},
\texttt{common\_invalid\_patterns},
\texttt{hint}, and \texttt{last\_updated\_generation}.

\item \textbf{Global crossover prior:}
\texttt{global\_transferable\_patterns},
\texttt{global\_harmful\_patterns},
\texttt{global\_complementarity\_patterns},
\texttt{hint}, and \texttt{last\_updated\_generation}.
\end{itemize}

Duplicate retries reuse the original generation context and append
the previous duplicate expression. Mutation retries may also
include recent deterministic validation failures. Expression
validity, DSL compliance, executability, and prior-schema
validation are handled deterministically rather than through a
separate LLM repair prompt.

\end{document}